\documentclass[journal]{IEEEtran}

\usepackage{amsmath,graphicx}
\usepackage{amssymb}
\usepackage{algpseudocode}
\usepackage{cite}
\usepackage{amsfonts}
\usepackage{graphicx,subfigure}
\usepackage{fancyhdr}  %
\usepackage{cases}
\usepackage{extarrows}
\usepackage{multirow,tabularx}
\usepackage{xcolor}

\begin{document}

\title{Reconfigurable Intelligent Surfaces and Metamaterials: The Potential of Wave Propagation Control 
for 6G Wireless Communications}
\author{George C. Alexandropoulos,~\IEEEmembership{Senior Member,~IEEE,} Geoffroy Lerosey,\\ M\'{e}rouane~Debbah,~\IEEEmembership{Fellow,~IEEE,} and Mathias~Fink
\thanks{G.~C.~Alexandropoulos is with the Department of Informatics and Telecommunications, National and Kapodistrian University of Athens, Panepistimiopolis Ilissia, 15784 Athens, Greece. (e-mail: alexandg@di.uoa.gr)}
\thanks{G. Lerosey and M. Fink are with Greenerwave, Incubator ESPCI, 75005 Paris, France. M. Fink is also with the Institut Langevin, ESPCI, 75005 Paris, France. (e-mails: \{geoffroy.lerosey, mathias.fink\}@espci.fr)}
\thanks{M. Debbah is with both the CentraleSup\'elec, University Paris-Saclay, 91192 Gif-sur-Yvette, France and the Mathematical and Algorithmic Sciences Lab, Paris Research Center, Huawei Technologies France SASU, 92100 Boulogne-Billancourt, France. (email: merouane.debbah@huawei.com)}
}

\maketitle

\begin{abstract}
The future 6th Generation (6G) of wireless communication networks will have to meet multiple requirements (e.g., throughput, latency, positioning accuracy, energy efficiency, massive connectivity, reliability, and networking  intelligence) in increasingly demanding levels, either individually or in combinations in small groups. This trend has spurred recent research activities on transceiver hardware architectures and novel wireless connectivity concepts. Among the emerging wireless hardware architectures belong the Reconfigurable Intelligent Surfaces (RISs), which are artificial planar structures with integrated electronic circuits that can be programmed to manipulate an incoming ElectroMagnetic (EM) field in a wide variety of functionalities. Incorporating RISs in wireless networks has been recently advocated as a revolutionary means to transform any naturally passive wireless communication environment to an active one. This can be accomplished by deploying cost-effective and easy to coat RISs to the environment’s objects (e.g., building facades and indoor walls/ceilings), thus, offering increased environmental intelligence for the scope of diverse wireless networking objectives. In this paper, we first provide a brief history on wave propagation control for optics and acoustics, and overview two representative indoor wireless trials at $2.47$GHz for spatial EM modulation with a passive discrete RIS. The first trial dating back to $2014$ showcases the feasibility of highly accurate spatiotemporal focusing and nulling, while the second very recent one demonstrates that passive RISs can enrich multipath scattering, thus, enabling throughput boosted communication links. Motivated by the late research excitement on the RIS potential for intelligent EM wave propagation modulation, we describe the status on RIS hardware architectures and present key open challenges and future research directions for RIS design and RIS-empowered 6G wireless communications.
\end{abstract}


\section{Introduction}
The increasingly demanding objectives for 6th Generation (6G) wireless communications have spurred recent research activities on novel wireless hardware architectures and connectivity concepts \cite{Saad}. The transceiver hardware architectures mainly include massive amounts of antennas or other electromagnetically excited elements, whose implementation cost and power consumption is extensively improved compared to conventional massive Multiple Input and Multiple Output (MIMO) systems \cite{mMIMO}. Hence, their role in 6G networks has the potential to be prominent via providing improved throughput, coverage extension, security, and positioning with lower power consumption footprint. In achieving these overarching goals, novel approaches for manipulating the wireless propagation signals and revolutionary networking schemes are required. To this end, there have been recently increased research interests (in most of the flagship magazines and conferences of the ComSoc as well as the Signal Processing and Antennas and Propagation Societies) in wireless connectivity concepts and signal processing algorithms incorporating nonconventional transceiver and ElectroMagnetic (EM) wave control architectures, like:
\begin{itemize}
\item Load modulated arrays and electronically steerable antenna radiators (e.g., \cite{Khandani,ESPARs,LMAs}); 
\item Hybrid Analog and Digital (A/D) beamforming structures (e.g., \cite{HBF,JSTSP2019,Yon2019}); and
\item Reconfigurable Intelligent Surfaces (RISs) (e.g., reflectarrays \cite{Hum2014}, metasurfaces \cite{metasurface1,metasurface2,metasurface3}, and antennas made from metamaterials \cite{smith1}).
\end{itemize}
The common philosophy of the works falling into this research direction is that: the more antennas deployed per transceiver or the more elements per electromagnetically excited structure, the more dominating hardware-imposed limitations will be present to the overall system design. These stricter and stricter limitations have to be tracked by sophisticated methods of channel modeling and compensated by intelligent signal/information processing algorithms and wireless connectivity techniques, including advanced machine learning tools and dedicated Artificial Intelligence (AI) methods with affordable computational complexity.

Over the last few years, metamaterials have emerged as a powerful technology with a broad range of applications, including wireless communications. Metamaterials comprise a class of artificial materials whose physical properties, and particularly their permittivity and permeability, can be engineered to exhibit various desired characteristics \cite{smith2}. When deployed in planar structures (a.k.a. metasurfaces), their effective parameters can be tailored to realize a desired transformation on the transmitted, received, or impinging EM waves \cite{smith3}. Such structures have been lately envisioned as a revolutionary means to transform any naturally passive wireless communication environment (the set of objects between a transmitter and a receiver constitute the wireless environment) to an active one \cite{Liaskos,RIS_TWC_2019,Marco_2019}. Their extremely low hardware footprint enables their cost-effective embedding in various components of the wireless propagation environment (e.g., building facades and room walls/ceilings), thus, enabling manmade EM wave propagation control and environmental AI. Due to latter reasons, RIS-empowered wireless communications are lately gaining booming attention for the upcoming 6G broadband networks.

In this paper, we commence in Section II with a brief history on wave propagation control, which has been a well-known concept in optics and acoustics. In Section III, we describe two indoor wireless trials at the WiFi frequency band for EM wave control with a fabricated passive RIS. The state-of-the-art in RIS hardware architectures for wireless communications is overviewed in in Section IV. Section V includes key open challenges and future research directions for RIS design and RIS-empowered wireless communication. The paper is concluded in Section VI. 

\section{Brief History of Wwave Propagation Control}
The control of propagation waves has been of significant interest in many domains ranging from medical imaging and therapy to wireless communications and nanolithography. Controlling waves in homogeneous media such as air is relatively easy and has long been realized using lenses in optics \cite{Goodman}. These apply a path difference to every ray going from one point to another, in order to allow for constructive wave interference at a specific 3-dimensional location, thus achieving highly accurate spatial signal focusing. Similarly, in microwave frequencies, reflectarrays have been proposed in order to steer EM waves to specific spatial directions \cite{Hum2014}. In both optics and microwave, the same principle has been applied. The waves are reflected from a planar matrix of resonators of different sizes, which applies a phase shift to the incoming wave that depends on the physical dimensions of its resonators. Reflectarrays are widely used for satellite communications and are the ancestors of the concept of metasurfaces (i.e., planar panels of metamaterials for wave control). They are extensively studied nowadays for free space applications in microwave, acoustics, and optics \cite{Science_2011}.

In heterogeneous and complex media, the control of wave propagation becomes more complex due to scattering and diffraction that can turn a plane wave into a completely random wave field. Those media have been considered, up to some years ago, as extremely difficult to tackle by wave physicists. It was shown, however, nearly two decades ago that those seemingly useless media can be tamed and used for the profits of highly accurate spatial focusing or imaging purposes through the concept of time reversal.  Time reversal constituted the broadband equivalent of phase conjugation, enabling scattering and reverberation harnessing in order tofocus waves far below the Rayleigh limit, which is given by the transmitting source aperture in free space \cite{Fink_TR_97,Draeger,Derode}. The time reversal technique has been also associated with locally resonant metamaterials in \cite{Lerosey_11,Lemoult1,Lemoult2} permitting wave focusing from the far field, way below the diffraction limit.

In optics, experiments with Spatial Light Modulators (SLMs) showed the possibility to focus light in media exhibiting multiple media (e.g. in \cite{Wang_2012}); this concept has been used in \cite{Katz} for imaging applications. SLMs consist of matrices of micro-mirrors or liquid crystal cells, which impose a physical phase shift to the portion of light they reflect. A simple incoherent energy-based feedback technique together with an optimization algorithm were proposed in \cite{Vellekoop} to focus light on a single speckle grain (i.e., a random wave field) through a thick layer of commercial paint. The idea in that work was to control the phase and/or amplitude of independent speckle grains at the input of a multiple scattering medium in order to add them in phase at its output, thus obtaining a focal spot whose intensity varies linearly with the number of controlled grains.

The aforementioned ideas of wave phase/amplitude control with reflectarrays, metamaterials, and SLMs have recently inspired the concept of RISs, as a revolutionary means for real-time reconfiguration of EM wave propagation in wireless communications. We next overview the first passive RIS (reflectarray) hardware architecture acting as a spatial EM modulator at $2.47$GHz \cite{metasurface2}, and its experimentation results for spatiotemporal focusing/nulling and multipath scattering enrichment in indoor room settings. 

\section{Indoor RIS-Empowered Wireless Trials}
In this section, we describe the design of an RIS structure (reflectarray) with nearly passive discrete elements, as presented in the seminal work of \cite{metasurface2} dating back to $2014$, and discuss its two representative indoor wireless trials at the WiFi frequency band.

\subsection{RIS Design}
Each unit cell element in the RIS design of \cite{metasurface1} was designed as a planar resonator intended to reflect the impinging EM waves with a controllable phase shift. In particular, each resonator was fabricated as a rectangular patch sitting on a ground plane and having two distinct states, resulting in binary phase modulation. According to these two states, the resonator reflects the waves either positively or negatively. The two resonator states were realized as follows. Suppose a resonance frequency $f_{\rm ref}$ which can be shifted using an electronic circuit. If $f_{\rm ref}$ is set such that it corresponds to the working frequency $f_0$, the resonant unit cell reflects the waves at this frequency with a $\pi$ phase shift compared to the bare ground plane. When $f_{\rm ref}$ is shifted away from $f_0$, the unit cell is non-resonant at this frequency, and the ground plane reflects the EM waves with a $0$ relative phase shift. It is noted that the phase shift of the reflected waves was defined relatively to that of the non-resonant unit cell, since it is general and can be applied to any kind of unit cell.
\begin{figure}[!t]
	\begin{center}
	\includegraphics[width=0.9\columnwidth]{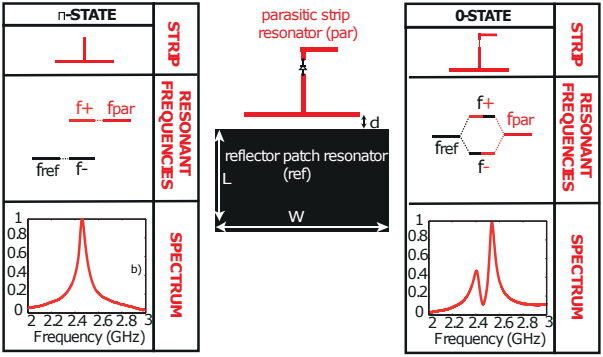}
	\caption{An RIS unit cell with binary phase tuning. A patch is sitting on a ground plane and is strongly coupled to a parasitic resonator that can be controlled with a PIN diode and a power feed \cite{metasurface1}. When the two resonators hybridize (diode forward mode), the resonator reflects the waves with a $0$ phase shift, while the main resonator only reflects microwaves with a $\pi$ phase shift (diode reverse mode).}
	\label{fig:Fig1}
	\end{center}
\end{figure} 

The schematic diagram of the designed RIS unit cell comprising of two strongly couple (or hybridized) resonators, and consuming only $50$mW, is illustrated in Fig$.$~\ref{fig:Fig1}. The first resonator is called the reflector patch resonator. It is a patch sitting on a ground plane, polarized along its short axis, and whose resonance frequency $f_{\rm ref}$ is set to $f_0$. The second resonator is the parasitic one being a strip line sitting on the ground plane and coupled to the reflector in the near field. Its resonance frequency $f_{\rm par}$ can be electronically tuned from $f_0$ to a higher frequency $f_1$ using a diode. When fpar is set to $f_1$, the reflector resonance frequency $f_{\rm ref}$ is unchanged and it reflects the waves with a $\pi$ phase shift compared to the bare ground plane (this is the $\pi$-state). In contrast, when the $f_{\rm par}$ is shifted to $f_0$, the two resonators hybridize and a dimer presenting two resonant frequencies $f-$ and $f+$, respectively below and above $f_0$, is created. In this state (i.e., at $f_0$), the dimer is again transparent and the waves are reflected by the ground plane with a $0$ relative phase shift (this is the $0$ state). This design presents notable advantages. The reflection properties of each RIS unit cell are insensitive to both the losses and impedance variations of the electronic components and to the soldering, which are only placed on the parasitic resonator.
\begin{figure}[!t]
	\begin{center}
	\includegraphics[width=0.9\columnwidth]{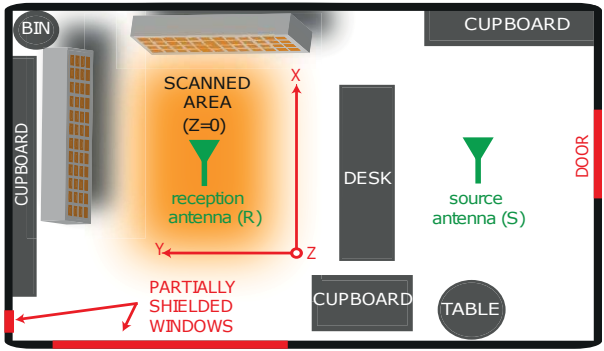}
	\caption{The RIS-empowered indoor wireless experimental setup of \cite{metasurface2} for highly accurate spatial focusing and nulling.}
	\label{fig:Fig2}
	\end{center}
\end{figure} 

\begin{figure}[!t]
	\begin{center}
	\includegraphics[width=0.9\columnwidth]{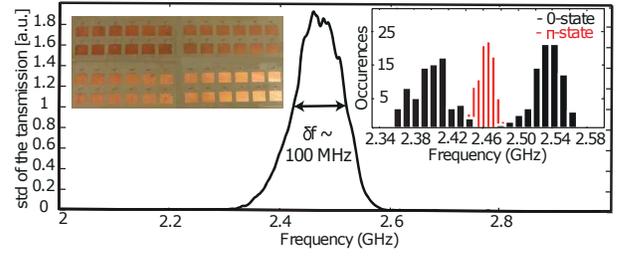}
	\caption{Standard deviation of the communication between the transmit (S) and receive (R) antennas for $11000$ random configurations and $10$ S positions, and one RIS panel (the inset left in Fig$.$~\ref{fig:Fig2}), as well as the distribution of the resonance frequencies of the $102$ resonators for both states measured with near field probes (inset right) \cite{metasurface2}. A portion of the fabricated $102$-element RIS is also illustrated.}
	\label{fig:Fig3}
	\end{center}
\end{figure} 

\subsection{Spatiotemporal Focusing and Nulling}
\begin{figure}[!t]
	\begin{center}
	\includegraphics[width=0.9\columnwidth]{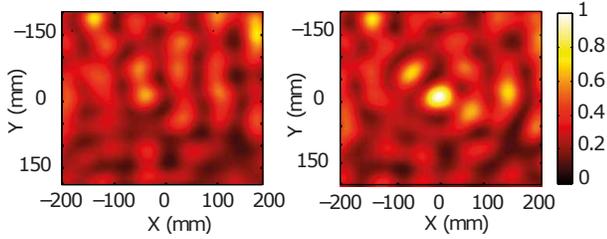}
	\caption{Without RIS (left) and with RIS (right) measured EM field intensity maps in the XY-plane averaged over $30$ realizations of the experiment for the spatiotemporal focusing use case \cite{metasurface2}. The R antenna is placed at the $(0,0)$ point.}
	\label{fig:Fig4}
	\end{center}
\end{figure}

\begin{figure}[!t]
	\begin{center}
	\includegraphics[width=0.9\columnwidth]{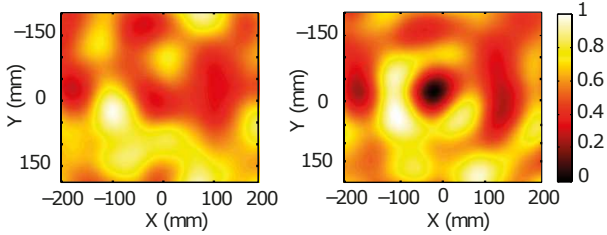}
	\caption{Without RIS (left) and with RIS (right) measured EM field intensity maps in the XY-plane averaged over 30 realizations of the experiment for the spatiotemporal nulling use case \cite{metasurface2}. The R antenna is placed at the (0,0) point.}
	\label{fig:Fig5}
	\end{center}
\end{figure}

The fabricated $0.4$m$^2$ and $1.5$mm of thickness RIS in \cite{metasurface2} consisted of $102$ controllable EM reflectors spaced by half a wavelength at the working frequency of $2.47$GHz (i.e., the distance between adjacent unit cells is $6$cm). The experimental setup for RIS-empowered indoor wireless communication is sketched in Fig$.$~\ref{fig:Fig1}. Figure~\ref{fig:Fig3} illustrates a portion of the designed RIS together with the standard deviation of the transmission between the transmit Source (S) and Receive (R) antennas for $11000$ random configurations, $10$ positions of the S antenna, and one RIS panel (the inset left in Fig$.$~\ref{fig:Fig1}). In this figure, the distribution of the resonance frequencies of the $102$ resonators for both states measured with near field probes (inset right) is also depicted. All RIS elements were controlled through two $54$-channels Arduino digital controllers, and an Agilent network analyzer was used to monitor the communication between the S and R antennas. Commercial monopole WiFi antennas polarized along the same axis were used as the RIS resonators. The S antenna was placed far away and out of the Line-Of-Sight (LOS) of R and the RIS in a furnished, and hence, rich scattering office room of dimensions $3$m$\times$$3$m$\times$$4$m. The R antenna was placed $1$m away from RIS. The conducted experimental results have showcased that the fabricated RIS can improve spatial focusing of the radiated EM wave onto well designed $\lambda/2$ isotropic focal spots (see Fig$.$~\ref{fig:Fig4}), or can be alternatively used for minimizing the EM field on the receiving R antenna placed at any location inside the room (see Fig$.$~\ref{fig:Fig5}). Interestingly and contrary to free space, it was shown that RIS results in an isotropic shaped EM field around the receiving antenna, which is attributed to the reverberant nature of the wireless propagation medium.

\subsection{Multipath Scattering Enrichment}
The fabricated RIS design of \cite{metasurface2}, and specifically a $65$-element portion of it, has been very recently deployed in \cite{metasurface3} in a $1.45$m$\times$$1$m$\times$$0.75$m chaotic aluminium cavity of volume $1.1$m$^3$, as shown in Fig. 6. The RIS covered the $4\%$ of the cavity’s surfaces and was deployed in order to boost the rate performance of a LOS link between two MIMO nodes each equipped with $8$ antenna elements. Identical commercial WiFi monopole antennas separated by $10$cm (a little more than $\lambda/2$ at the working frequency of $2.47$GHz) were used at both antenna arrays. All antennas were in the same orientation (i.e., no polarization diversity applied), and each antenna array was connected to a Radio Frequency (RF) switch, which in turn was connected to a vector network analyzer.
The 65 RIS unit cells in the experimental setup of Fig. 6 were designed according to the iterative sequential optimization algorithm in \cite{Vellekoop2} in order to improve the rank of the $N\times N$ MIMO channel propagation matrix with $2\leq N\leq8$,compared to the case where the RIS is not used. For each iteration of the deployed algorithm, the binary configuration ($0$ and $\pi$ phase states) of a unit cell was changed, the new channel matrix was measured, and the new effective channel rank was calculated. If that phase change resulted in higher channel rank, the unit cell's configuration was updated accordingly. With this iterative way, the configurations of all unit cells were updated. It is noted that the configuration of each unit cell was iterated multiple times to deal with the long-range correlations between the RIS optimal configurations, due to the reverberation inside the cavity. The obtained experimental results are demonstrated in Fig$.$~\ref{fig:Fig7} for $N=2$, $4$, and $6$. As shown for all considered values for $N$, the optimized RIS configuration leads to a full rank channel as the orthogonal one, which models the richest multipath scattering conditions. It can be also seen that the convergence to the highest possible channel rank depends on the value of $N$.
\begin{figure}[!t]
	\begin{center}
	\includegraphics[width=0.8\columnwidth]{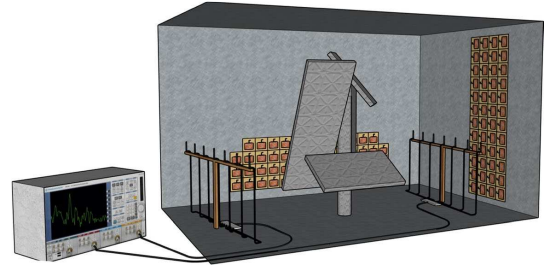}
	\caption{The aluminium cavity experimental setup of \cite{metasurface3} incorporating \cite{metasurface2}'s fabricated RIS coated on the wall in order to enrich multipath scattering in a LOS $N\times N$ MIMO wireless communication link at the WiFi frequency 2.47GHz.}
	\label{fig:Fig6}
	\end{center}
\end{figure} 

\begin{figure}[!t]
	\begin{center}
	\includegraphics[width=0.75\columnwidth]{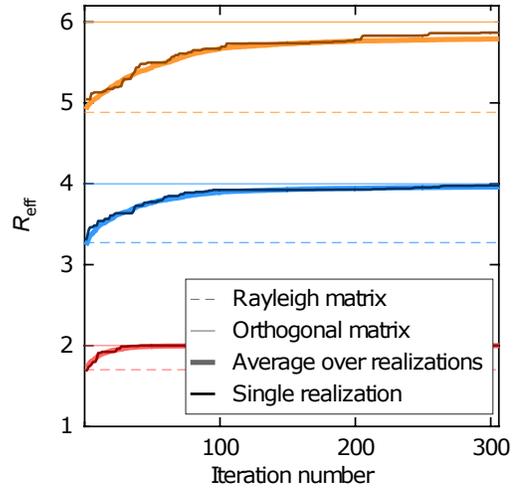}
	\caption{The evolution of the effective rank ($R_{\rm eff}$) of the LOS $N\times N$ channel matrix (red for $N=2$, blue for $N=4$, and orange for $N=6$) over the number of algorithmic steps for a single channel realization and for the average over $30$ channel realizations with different MIMO channel geometries. Benchmarks for Rayleigh fading and perfect channel orthogonality are also included.}
	\label{fig:Fig7}
	\end{center}
\end{figure} 

It is obvious from Fig$.$~\ref{fig:Fig7} that the optimal channel diversity can be achieved by physically shaping the propagation medium itself via \cite{metasurface2}'s fabricated RIS design having nearly passive unit cells. The disorder of the random EM propagation environment inside the cavity was actually tamed to impose perfect orthogonality of the wireless channels. As further demonstrated in \cite{metasurface3} through an indoor wireless image transmission trial using a LOS $3\times3$ MIMO link empowered by the fabricated RIS, the number of effective independent propagation channels reached the maximum number $3$, which was only equal to $2$ when the RIS was not used. This fact was translated to improvement in the achievable rate performance.

\section{State-of-the-Art In \\RIS Hardware Architectures}
In this section, we overview the late advances in RIS hardware architecture designs for wireless communications based on three different categorizations \cite{HMIMOS}. The first category focuses on whether an RIS includes active or passive components, which consequently determines its overall power consumption. The second category is based on whether an RIS is a contiguous surface or is comprised of discrete elements, and the third category discusses the available RIS modes of operation when deployed in wireless communication systems.

\subsection{RIS Power Consumption}
\subsubsection{Active RISs} To realize reconfigurable wireless environments, an RIS can serve as a transmitter, receiver, or an EM wave modulator. When the transceiver role is considered, and thus energy-intensive RF circuits and signal processing units are embedded in the surface, the term active RIS is adopted \cite{Yurduseven,HMIMOS,Hu2018, Pizzo, Nir2019, Taha, Mine_ICASSP2020}. Many RF chains realizing a special form of hybrid A/D beamforming have been considered in \cite{Nir2019} (currently, only for transmission), whereas [36] proposes that all RIS unit elements are attached to a single RF chain to enable efficient wireless channel estimation at the RIS side. On another note, active RIS systems comprise a natural evolution of conventional massive MIMO systems, by packing more and more software-controlled antenna elements onto a 2-dimensional surface of finite size. In \cite{Hu2018}, where the spacing between the RIS unit elements reduces when their number increases, an active RIS is also termed as large intelligent surface. A practical implementation of active RISs can be a compact integration of a large number of tiny antenna elements with reconfigurable processing networks realizing a continuous antenna aperture. This structure can be used to transmit and receive communication signals across the entire surface by leveraging the hologram principle \cite{Yurduseven,Pizzo}. Another active RIS implementation can be based on discrete photonic antenna arrays that integrate active optical-electrical detectors, converters, and modulators for performing transmission, reception, and conversion of optical or RF signals \cite{Yurduseven}.

\subsubsection{Passive RISs} As previously discussed, a passive RIS acts like a passive metal mirror or wave modulator, and can be programmed to change an impinging EM field in a customizable way \cite{metasurface2,RIS_TWC_2019,Marco_2019}. Compared with its active counterpart, a passive RIS is usually composed of low cost passive elements that do not require dedicated power sources. Their circuitry and embedded sensors can be powered with energy harvesting modules, an approach that has the potential of making them truly energy neutral. Regardless of their specific implementations, what makes the passive RIS technology attractive from an energy efficiency standpoint, is its inherent capability in controlling waves impinging upon it, and forwarding the incoming signal without employing any power amplifier nor RF chain, and also without applying sophisticated signal processing techniqes. Moreover, passive RISs can work in full duplex mode without significant self-interference or increased noise level, and require only low rate control link or backhaul connections. Finally, passive RIS structures can be easily integrated into the wireless communication environment, since their extremely low power consumption and hardware costs allow them to be deployed into building facades, room and factory ceilings, laptop cases, or even human clothing \cite{RIS_TWC_2019,Marco_2019}.

\subsection{RIS Hardware Structures}
\subsubsection{Contiguous RISs} A contiguous RIS integrates a virtually infinite number of elements into a limited surface area in order to form a spatially continuous transceiver aperture \cite{Yurduseven,Pizzo}. For a better understanding of the operation of contiguous surfaces and their communication models, we commence with a brief description of the physical operation of the optical holography concept. Holography is a technique that enables an EM field, which is generally the result of a signal source scattered off objects, to be recorded based on the interference principle of the EM propagation wave. The recorded EM field can be then utilized for reconstructing the initial field based on the diffraction principle. It should be noted that wireless communications over a continuous aperture is inspired by optical holography. Since a continuous aperture benefits from the theoretical integration of an infinite number of antennas which can be viewed as the asymptotic limit of Massive MIMO, its potential advantages include the highly accurate spatial resolution, and the creation/ detection of EM waves with arbitrary spatial frequency components, without undesired side lobes.

\subsubsection{Discrete RISs} A discrete RIS is usually composed of many discrete unit cells made of low power software-tunable metamaterials \cite{metasurface2,RIS_TWC_2019,Marco_2019,Yurduseven,Hu2018,Pizzo,Nir2019}. The means to electronically modify the EM properties of the unit cells range from off the shelves electronic components to using liquid crystals, MicroElectroMechanical Systems (MEMS) or even electromechanical switches, and other reconfigurable metamaterials. This structure is substantially different from the conventional multi-antenna antenna array. One embodiment of a discrete surface is based on discrete ‘meta-atoms’ with electronically steerable reflection properties \cite{Liaskos}. As mentioned earlier, another type of discrete surface is the active one based on photonic antenna arrays. Compared with contiguous RISs, discrete RISs have some essential differences from the perspectives of implementation and hardware \cite{HMIMOS}. Current RIS fabricated designs comprise of discrete unit elements.

\subsection{RIS Operation Modes}
The following four RIS operating modes are lately considered in wireless communications: 1) continuous RISs acting as active transceivers; 2) discrete RISs used also as active transceivers; 3) discrete RISs with a single RF chain intended for wireless channel estimation; and 4) discrete passive RISs operating as EM wave modulators. 

\subsubsection{Continuous RISs as Active Transceivers} According to this mode of operation, a continuous RIS operates as an active transceiver. The RF signal is generated at its backside and propagates through a steerable distribution network to the contiguous surface constituted by a large number of software-defined and electronically steerable elements that generate multiple beams to the intended users. A distinct difference between and active continuous RIS and a passive one is that the beamforming process of the former is accomplished based on the holographic concept, which is a new dynamic beamforming technique based on software-defined EM wave modulators with low cost/weight, compact size, and a low-power circuitry. 

\subsubsection{Discrete RISs as Active Transceivers} Discrete active RISs, also known as Dynamic Metasurface Antennas (DMAs), have been recently proposed as an efficient realization of massive antenna arrays for wireless communications \cite{smith1,Nir2019}. They provide beam tailoring capabilities using simplified transceiver hardware, which requires less power and cost compared to conventional hybrid A/D/ antenna arrays (i.e., those based on patch arrays and phase shifters), eliminating the need for complicated corporate feed as well as active phase shifters. DMAs may comprise of a large number of tunable metamaterial antenna elements that can be packed in small physical areas \cite{Vellekoop2} for a wide range of operating frequencies. This feature makes them an appealing technology for the next generation extreme massive MIMO transceivers.

\subsubsection{Discrete RISs for Channel Estimators} An RIS architecture comprising of any number of passive reflecting elements, a simple controller for their adjustable configuration, and a single RF chain for baseband measurements was presented in [36]. Capitalizing on this architecture and assuming sparse wireless channels in the beamspace domain, an alternating optimization approach for explicit estimation at the RIS side of the channel gains at its elements (which are all attached to the single RF chain) was proposed.

\subsubsection{Discrete RISs as EM Wave Modulators:} Another operation mode of RISs is the mirror or wave modulator, where the surface is considered to be discrete and passive. In this case, an RIS includes unit cells that can reconfigured in real time (as \cite{metasurface2}'s reflectarray that was previously described), which makes their synergetic operation resembling that of spatiotemporal focusing. \cite{metasurface1}, unlike the late considerations for continuous transceiver RIS systems. It is worth noting that most of the existing works (e.g., \cite{LMAs,JSTSP2019,Yon2019}) focus on this RIS operation mode which is simpler to implement and analyze.

\section{Open Challenges and Future Directions}
As highlighted in the previous sections and as witnessed from the recent RIS-focused demonstrations and increasingly booming research interests, RISs have high potential in enabling smart wireless propagation environments via their delicate EM wave control capabilities. In addition, their hardware design principles can pave the way for the fabrication of the next generation ultra massive MIMO antenna systems with acceptable power consumption levels and affordable hardware complexity. Although, RIS-enabled wave control dates back to acoustics, recent experimental results and hardware designs showcased the dynamics of the various RIS architectures (especially those stemming from metamaterials) for higher frequency communications (from sub6GHz to THz), which will be a core component of 6G wireless communication networks.

Owing to the nearly zero power consumption of passive RISs and their envisioned extremely low fabrication cost and compact size, the ubiquitous deployment (i.e., ultra densification) of intelligent EM wave control structures becomes feasible, which will eventually lead to the realization of the holographic concept \cite{HMIMOS,Pizzo} for 6G wireless communication. This concept will enable highly accurate multi-spot spatiotemporal focusing for diverse communication objectives (e.g., increase highly localized throughput, accurate positioning, improved security, and reduced EM field exposure) with minimal realization overhead. It is also noted that holographic wireless systems can be further reinforced with active RISs made from power efficient metamaterials. However, to unveil the true potential of holographic communications and devise efficient relevant algorithmic approaches, synergies among the theorists and practitioners in the fields of EM propagation, antenna design, and wireless communications are required. Luckily, and in contrast to the unfortunate parallel research paths up to date, the RIS research topic has pronounced the common ground of the latter fields witnessing the increased needs for higher convergence between electromagnetics, communication, and wireless systems’ theories.

In the following, we emphasize on some of the key open challenges and future research directions (see also \cite{HMIMOS,NirDMA}) with the fabrication of individual RIS hardware structures as well as the design of wireless networking schemes empowered by RISs as EM wave controllers.

\subsubsection{\textbf{Electromagnetics Information Theory}} It is natural to expect that RIS-empowered wireless communication systems will exhibit different features compared with traditional communications based on conventional multi-antenna transceivers. Recall that current communication systems operate over uncontrollable wireless environments, whereas wireless systems incorporating EM wave modulating RISs will be capable of reconfiguring signal propagation. This fact witnesses the need for new information theoretical methodologies to characterize the physical channels in RIS-empowered systems and analyze their ultimate capacity gains (e.g., study optimal signaling and multi-user communcation), as well as for new signal processing algorithms and networking schemes for realizing wireless communications via RISs. For example, continuous RISs can be used for the reception and transmission of the impinging EM field over its continuous aperture using the hologram concept. Differently from the massive MIMO systems, RIS operation can be described by the Fresnel-Kirchhoff integral that is based on the Huygens-Fresnel principle \cite{metasurface2,Yurduseven}. It also interesting to devise and analyze schemes based on time reversal (e.g., \cite{Draeger,Lerosey_11,Mine_TR}) in the context of future large-bandwidth communications empowered by RISs (especially in the millimeter and THz bands), in order to enable highly accurate multi-spot spatiotemporal focusing for diverse communication objectives. To this end, the implications of the RIS role over different frequency bands need to be further investigated. For example, at very high frequencies, the EM field is not as diffuse in nature as it is in low frequencies. This entails RISs to be deployed as highly localized access point extenders, rather than structures that are capable to enrich multipath scattering (as has been done for throughput boosting in WiFi frequencies).

\subsubsection{\textbf{Modeling of RIS-Empowered EM Wave Propagation}} Realistic models for the EM wave propagation of signals bouncing on RISs are needed. Additionally, the adoption of large RISs challenges the common far-field EM propagation assumption. The sources of information signals can be close to RISs, and particularly, in distances smaller than the RIS structure size, giving rise to near-field EM propagation. It is thus of paramount importance to devise physics-inspired models for EM wave propagation in the RIS proximity, which account for interactions in the RIS circuitry. It also necessary to study realistic pathloss models for RIS-empowered wireless networks (especially for the passive RIS case), in order to unveil the RISs’ true link budget potential as well as their optimal placement in space. Moreover, the interactions of the RIS unit elements, which are placed in subwavelength distances in RISs made from metamaterials, need to accurately modeled and incorporated in the channel matrix model in order to be accounted for in the communication theory analyses and wireless signal processing designs.

\subsubsection{\textbf{Channel Tracking in RIS-Empowered Networks}} To date, studies on RIS-empowered wireless communications assume that the transmitters/receivers have full channel information knowledge. In practice, however, the channel coefficients need to be efficiently estimated, which is a challenging task with either passive or active RISs. Channel estimation cannot be implemented at the side of a passive RIS, but rather at one end (transmitter or receiver) of the communication link. This makes the channel estimation task challenging and has motivated the introduction of channel estimation protocols for the case of passive-RIS-empowered wireless communications. However, current approaches require lengthy channel estimation protocols, and low overhead channel estimation frameworks are needed. When channel estimation is carried out in a time-division duplexing manner, active RISs offer the possibility of tuning their elements to facilitate channel estimation via pilot signals, and to adapt in a manner which optimizes data reception in light of the estimated channel. The design and analysis of efficient algorithms for active RISs, which have to estimate features of the wireless channel and reliably communicate, have not yet been properly treated.

\subsubsection{\textbf{Algorithms for RIS-Enabled EM Wave Control}} Channel dependent beamforming has been extensively considered in massive MIMO systems. However, realizing environment-aware designs in RIS-empowered wireless communication systems is extremely challenging, since the RIS unit cells, which can be fabricated from metamaterials, impose demanding online tuning constraints. The latest RIS design formulations include large numbers of reconfigurable parameters with nonconvex constraints, rendering their optimal solution highly nontrivial. For the case of continuous RISs, intelligent holographic beamforming is an approach to smartly target and follow individual or small clusters of devices, and provide them with high fidelity beams and smart radio management. However, self-optimizing holographic beamforming technologies that depend on complex aperture synthesis and low level modulation are not available yet.

Current algorithmic designs for active RISs focus either on narrowband communications or ignore their capability to dynamically configure the frequency-selective profile of each unit metamaterial element. This unique property, which does not exist in any conventional hybrid A/D architecture, provides increased flexibility for the design of wideband operation by matching the spectral behavior of each element to optimize the equivalent wideband channel. Consequently, the true potential of extreme massive MIMO systems implemented with active RISs in achieving ultra-reliable and ultra-high rate communications is not yet fully explored \cite{NirDMA}.

\subsubsection{\textbf{Design of Passive and Active RIS Architectures}} A large body of fabricated designs and experimental works is still required in order to transit the RIS concept into an established technology for 6G wireless communications. As previously discussed, densely deployed EM wave modulating RISs have the potential to enable massive numbers of highly focused beams for various communication objectives (e.g, massive data streams for multiple spatial spots when throughput is the objective). In addition, the future designs need to address the provisioned requirements for the millimeter and the THz bands. In such cases, efficient hardware designs are necessary, which currently quite challenging.

Since both active and passive RISs are lately gaining increased interest for both EM wave propagation control and transmission/reception, it is reasonable to envision hybrid passive and active RISs. Such structures will notably strengthen the design flexibility for RISs, either for enabling programmable wireless environments, or realizing ultra massive MIMO antenna arrays, or both. For instance, having such a hybrid RIS acting as a receiving device [36] can significantly facilitate channel estimation via machine learning tools \cite{JSTSP2019}, which is still a major challenge and a source of substantial communication control overhead in purely passive RISs. In addition, hybrid RISs will enable more advanced relaying strategies, overcoming the dominating impact of pathloss in the applications of their passive versions.

\subsubsection{\textbf{Use Cases for RIS-Empowered Connectivity}} The use cases and applications where passive and active RISs can provide substantial improvement compared to current transceiver and network architectures have not yet been thoroughly identified. For example, the RIS planar shape and compact size for the passive versions as well as the active versions with small numbers of RF chains, facilitate their deployment in indoor environments, like buildings, factories, malls, train stations, hospitals, and airports. In such setups, RISs are expected to communicate with multiple users in close to LOS conditions, possibly operating in the near-field regime. As previously discussed, such near-field scenarios bring forth the possibility of spatiotemporal focusing, namely, the ability to focus the signal towards a specific location in space, instead of a specific direction as in the far-field conditions via conventional beamforming. Moreover, the potential of passive and active RISs in outdoor network setups needs to be further investigated and demonstrated, and for different operating frequencies. Finally, a large body of works that combines RISs with various other communication technologies (e.g., physical layer security, unamanned aerial vehicles, energy harvesting, and cognitive networking) have lately appeared, identifying various relevant design challenges.

\subsubsection{\textbf{Joint Communication and Computing RIS Platforms}} Late advances in chipset design and computational effectiveness of AI approaches have enabled the incorporation of basic AI functionalities in 5th Generation (5G) base stations and mobile handsets. Following this trend and the basic computing and storage capabilities of the current RIS designs, we envision future RISs to being capable of training local Artificial Neural Networks (ANNs) to obtain models for their local wireless connectivity environment. Each ANN can operate on the unit cell level, where each cell trains a model and all derived models from the unit cells are used to design the global model for the RIS. This model can be used for efficient online configurations as per the desired EM wave control. Alternatively, each RIS model can be trained from all the unit cells simultaneously.

In achieving the latter overarching goal, further advances in low complexity AI approaches are needed. To this direction belong the binary neural networks, which are lately receiving significant attention for smart mobile handsets. These ANNs have binary weights and are activated at run time. At the training time, the weights and activations are used for computing gradients, however, the gradients and true weights are stored in full precision. This procedure permits effective ANN training on systems with limited resources. The availability of an individual ANN model per RIS structure can be used for future configurations of the values of all the deployed RISs in the RIS-empowered wireless network. The individual ANNs can be also shared to a central network entity that gathers in a compressed manner the sensing information from the available dense network of AI-enabled RISs. This sensing information can be used for network monitoring, management, and optimization purposes.

\section{Conclusions}
The concepts of artificial EM wave propagation control and tunable reflecting metamaterials, which naturally constitute the ancestors of smart programmable wireless environments and RISs, have been initially conceived in the acoustics and optics fields dating back to more than ten years ago. Due to the increased potential of RISs for 6G wireless communication networks, as witnessed by the recent proof of concepts with both passive reflectarrays and active metasurface antennas, there has been lately a surprisingly increasing attention on the RIS topic from both academia and industry working in antenna design and wireless communications. In fact, a large body of research papers and special issues in prestigious ComSoc and Antenna Propagation Society periodicals has appeared in the last 2 years, as well as novel RIS demonstrators and collaborative R\&D projects.

In this paper, we provided a brief historical description on wave propagation control for optics and acoustics, and overviewed two representative indoor wireless trials at $2.47$GHz for EM wave modulation with a fabricated passive discrete RIS. The first trial dating back to $2014$ showcased the feasibility of highly accurate spatiotemporal focusing and nulling, while the second very recent one in $2019$ demonstrated that passive RISs can enrich multipath scattering, thus, enabling throughput boosted wireless connectivity. We also discussed the current status in RIS hardware designs emphasizing the key features of the different approaches. We concluded the paper with a detailed list of key open challenges and future research directions for the design of individual RIS structures as well as for connectivity approaches in RIS-empowered wireless networks. As advocated in numerous parts of this paper, the RIS topic triggers fascinating synergies among the fields of EM propagation, antenna design, communication theory, and signal processing for wireless communications. More importantly, we believe that potential of RISs in EM wave propagation control will enable highly accurate multi-spot spatiotemporal focusing towards the ultimate goal for revolutionary 6G wireless communication networks with embedded environmental intelligence.



\begin{thebibliography}{99}
\bibitem{Saad} W. Saad, M. Bennis, and M. Chen, ``A vision of 6G wireless systems: Applications, trends, technologies, and open research problems,'' \textit{IEEE Network}, 2019.

\bibitem{mMIMO} F. Rusek, D. Persson, B. K. Lau, E. G. Larsson, T. L.Marzetta, O. Edfors, and F. Tufvesson, ``Scaling up MIMO: Opportunities and challenges with very large arrays,'' \textit{IEEE Signal Process. Mag.}, vol. 30, no. 1, pp. 40--60, Jan. 2013.

\bibitem{Khandani} A. K. Khandani, ``Media-based modulation: A new approach to wireless transmission,'' in \textit{Proc. IEEE ISIT}, Istanbul, Turkey, Jul. 2013, pp. 3050--3054.

\bibitem{ESPARs} G. C. Alexandropoulos, V. I. Barousis, and C. B. Papadias, ``Precoding for multiuser MIMO systems with single-fed parasitic antenna arrays,'' in \textit{Proc. IEEE GLOBECOM}, Austin, USA, Dec. 2014, pp. 3656--3661.

\bibitem{LMAs} M. A. Sedaghat, V. I. Barousis, R. R. M\"{u}ller, and C. B. Papadias, ``Load modulated arrays: A low-complexity antenna,'' \textit{IEEE Commun. Mag.}, vol. 54, no. 3, pp. 46--52, Mar. 2016.

\bibitem{HBF} A. F. Molisch, V. V. Ratnam, S. Han, Z. Li, S. L. H. Nguyen, L. Li, and K. Haneda, ``Hybrid beamforming for massive MIMO: A survey,'' \textit{IEEE Commun. Mag.}, vol. 55, no. 9, pp. 134--141, Sep. 2017.

\bibitem{JSTSP2019} E. Vlachos, G. C. Alexandropoulos, and J. Thompson, ``Wideband MIMO channel estimation for hybrid beamforming millimeter wave systems via random spatial sampling,'' \textit{IEEE J. Sel. Topics Signal Process.}, vol. 13, no. 5, pp. 1136--1150, Sep. 2019.

\bibitem{Yon2019} S. S. Ioushua and Y. C. Eldar, ``A family of hybrid analog–digital beamforming methods for massive MIMO systems,'' \textit{IEEE Trans. Signal Process}. vol. 67, no. 12, pp. 3243--3257, Jun. 2019.

\bibitem{Hum2014} S. V. Hum and  J. Perruisseau-Carrier, ``Reconfigurable reflectarrays and array lenses for dynamic antenna beam control: A review,'' IEEE Trans. Antennas Prop., vol. 62, no. 1, pp. 183--198, Jan. 2014.

\bibitem{metasurface1} H. Yang, X. Cao, F. Yang, J. Gao, S. Xu, M. Li, Z. Chen, Y. Zhao, Y. Zheng, and S. Li, ``A programmable metasurface with dynamic polarization, scattering and focusing control,'' \textit{Scientific Reports}, vol. 6, pp. 1--11, Oct. 2016.

\bibitem{metasurface2} N. Kaina, M. Dupre, G. Lerosey, and M. Fink, ``Shaping complex microwave fields in reverberating media with binary tunable metasurfaces,'' \textit{Scientific Reports}, vol. 4, pp. 1--7, Aug. 2014.

\bibitem{metasurface3} P. del Hougne, M. Fink, and G. Lerosey, ``Optimally diverse communication channels in disordered environments with tuned randomness,'' \textit{Nature Electronics}, vol. 2, pp. 36--41, Jan. 2019.

\bibitem{smith1}	I. Yoo, M. F. Imani, T. Sleasman, H. D. Pfister, and D. R. Smith, ``Enhancing capacity of spatial multiplexing systems using reconfigurable cavity-backed metasurface antennas in clustered MIMO channels,'' \textit{IEEE Trans. Commun.}, vol. 67, no. 2, pp. 1070--1084, Feb. 2018.

\bibitem{smith2} D. R. Smith, J. B. Pendry, and M. C. Wiltshire, ``Metamaterials and negative refractive index,'' \textit{Science}, vol. 305, no. 5685, pp. 788--792, 2004.

\bibitem{smith3} D. R. Smith, O. Yurduseven, L. Pulido-Mancera, P. Bowen, and N. B. Kundtz, ``Analysis of a waveguide-fed metasurface antenna,'' \textit{Physical Review Applied}, vol. 8, no. 5, Nov. 2017.

\bibitem{Liaskos} C. Liaskos, S. Nie, A. Tsioliaridou, A. Pitsillides, S. Ioannidis, and I. F. Akyildiz, ``A new wireless communication paradigm through software- controlled metasurfaces,'' \textit{IEEE Commun. Mag.}, vol.  56,  no. 9, pp. 162--169, Sep. 2018.

\bibitem{RIS_TWC_2019} Huang, A. Zappone, G. C. Alexandropoulos, M. Debbah, and C. Yuen, ``Reconfigurable intelligent surfaces for energy efficiency in wireless communication,'' \textit{IEEE Trans. Wireless Commun.}, vol. 18, no. 8, pp. 4157--4170, Aug. 2019.

\bibitem{Marco_2019} M. Di Renzo, M. Debbah, D.-T. Phan-Huy, A. Zappone, M.-S. Alouini, C. Yuen, V. Sciancalepore, G. C. Alexandropoulos, J. Hoydis, H. Gacanin, J. de Rosny, A. Bounceu, G. Lerosey, and M. Fink, ``Smart radio environments empowered by reconfigurable AI meta-surfaces: an idea whose time has come,'' EURASIP J. Wireless Commun. Netw., vol. 2019, no. 1, pp. 1--20, May 2019.

\bibitem{Goodman} J. W. Goodman, \textit{Introduction to Fourier optics}, Englewood, Roberts \& Company Publishers, 2005.

\bibitem{Science_2011} N. Yu, P. Genevet, M. A. Kats, F. Aieta, J.-P. Tetienne, F. Capasso, and Z. Gaburro, ``Light propagation with phase discontinuities: Generalized laws of reflection and refraction,'' \textit{Science}, vol. 334, no. 6054, pp. 333--337, 2011.

\bibitem{Fink_TR_97} M. Fink, ``Time reversed acoustics,'' \textit{Physics Today}, vol. 50, no. 3, Mar. 1997.

\bibitem{Draeger} C. Draeger, and M. Fink, ``One-channel time reversal of elastic waves in a chaotic 2D-silicon cavity,'' \textit{Physical Review Lett.}, vol. 79, Article ID 407, Jul. 1997.

\bibitem{Derode} A. Derode, A. Tourin, J. de Rosny, M. Tanter, S. Yon, and M. Fink, ``Taking advantage of multiple scattering to communicate with time-reversal antennas,'' \textit{Physical Review Lett.}, vol. 90, Article ID 014301, Jan. 2003.

\bibitem{Lerosey_11} G. Lerosey, J. de,Rosny, A. Tourin, and M. Fink, ``Focusing beyond the diffraction limit with far-field time reversal,'' \textit{Science}, vol. 315, no. 5815, pp. 11--20, Feb. 2007.

\bibitem{Lemoult1} F. Lemoult, M. Fink, and G. Lerosey, ``Acoustic resonators for far-field control of sound on a subwavelength scale,'' \textit{Physical Review Lett.}, vol. 107, Article ID 064301, Aug. 2011.

\bibitem{Lemoult2} F. Lemoult, G. Lerosey, J. de Rosny, and M. Fink, ``Resonant metalenses for breaking the diffraction barrier,'' \textit{Physical Review Lett.}, vol. 104, Article ID 203901, May 2010.

\bibitem{Wang_2012} Y. M. Wang, B. Judkewitz, C. A. Dimarzio, and C. Yang, ``Deep-tissue focal fluorescence imaging with digitally time-reversed ultrasound-encoded light,'' \textit{Nature Commun.}, vol. 3, no. 928, Jun. 2012.

\bibitem{Katz} O. Katz, E. Small, Y. Bromberg, and Y. Silberberg, ``Focusing and compression of ultrashort pulses through scattering media,'' \textit{Nature Photonics}, vol. 5, pp. 372--377, May 2011.

\bibitem{Vellekoop} I. M. Vellekoop and A. P. Mosk, ``Focusing coherent light through opaque strongly scattering media'' \textit{Optics Lett.}, vol. 32, no. 16, pp. 2309--2311, 2007.

\bibitem{HMIMOS} C. Huang, S. Hu, G. C. Alexandropoulos, A. Zappone, C. Yuen, R. Zhang, M. Di Renzo, and M. Debbah, ``Holographic MIMO surfaces for 6G wireless networks: Opportunities, challenges, and trends,'' \textit{IEEE Wireless Commun.}, accepted for publication, 2020.

\bibitem{Yurduseven} O. Yurduseven, D. L. Marks, T. Fromenteze, and D. R. Smith, ``Dynamically reconfigurable holographic metasurface aperture for a mills-cross monochromatic microwave camera,'' \textit{Optics Express}, vol. 26, no. 5, pp. 5281--5291, 2018.

\bibitem{Pizzo} A. Pizzo, T. L. Marzetta, and L. Sanguinetti, ``Spatially-stationary model for holographic MIMO small-scale fading,'' \textit{[online] https://arxiv.org/abs/1911.04853}, 2019.

\bibitem{Hu2018} S. Hu, F. Rusek, and O. Edfors, ``Beyond massive MIMO: The potential of positioning with large intelligent surfaces,'' \textit{IEEE Trans. Signal Process.}, vol. 66, no. 7, pp. 1761--1774, Apr. 2018.

\bibitem{Nir2019} N. Shlezinger, O. Dicker, Y. C. Eldar, I. Yoo, M. F. Imani, and D. R. Smith, ``Dynamic metasurface antennas for uplink massive MIMO systems,'' \textit{IEEE Trans. Commun.}, vol. 67, no. 10, pp. 6829--6843, Oct. 2019.

\bibitem{Taha} A. Taha, M. Alrabeiah, and A. Alkhateeb, ``Enabling large intelligent surfaces with compressive sensing and deep learning,'' \textit{[online] https://arxiv.org/abs/1904.10136v2}, 2019.

\bibitem{Mine_ICASSP2020} G. C. Alexandropoulos and E. Vlachos, ``A hardware architecture for reconfigurable intelligent surfaces with minimal active elements for explicit channel estimation,'' in \textit{Proc. IEEE ICASSP}, Barcelona, Spain, May 2020, pp. 9175--9179.

\bibitem{Akyildiz_THz} I. F. Akyildiz and J. M. Jornet, ``Realizing ultra-massive MIMO ($1024\times1024$) communication in the ($0.06$-$10$) terahertz band,” \textit{Nano Commun. Netw.}, vol. 8, pp. 46--54, 2016.

\bibitem{Vellekoop2} I. M. Vellekoop and A. P. Mosk, ``Phase control algorithms for focusing light through turbid media,'' \textit{Optics Commun.}, vol. 281, no. 11, pp. 3071--3080, 2008.

\bibitem{NirDMA} N. Shlezinger, G. C. Alexandropoulos, M. F. Imani, Y. C. Eldar, and D. R. Smith, ``Dynamic metasurfaces antennas for 6G massive MIMO communications,'' \textit{IEEE Commun. Mag.}, under review, \textit{[online] https://arxiv.org/pdf/2006.07838.pdf}, 2020.

\bibitem{Mine_TR} G. C. Alexandropoulos, R. Khayatzadeh, M. Kamoun, Y. Ganghua, and M. Debbah, ``Indoor time reversal wireless communication: Experimental results for localization and signal coverage,'' in \textit{Proc. IEEE ICASSP}, Brighton, UK, May 2019, pp. 1--5.
\end{thebibliography}
\end{document}